\documentclass[%
 reprint,
 showpacs,
 showkeys,
 altaffilletter,
 superscriptaddress,
 nobibnotes,
 longbibliography,
 amsmath,amssymb,
 aps,
 prapplied,
 floatfix,
]{revtex4-1}

\usepackage{dcolumn}
\usepackage{bm}
\usepackage{graphicx}
\usepackage{pslatex}
\usepackage{xspace}
\usepackage{subfigure}
\usepackage[utf8]{inputenc}
\usepackage[T1]{fontenc}
\usepackage{amsmath}
\usepackage{textcomp}
\usepackage{color}

\begin{document}

\preprint{APS/123-QED}

\title{Improved evaluation of deep-level transient spectroscopy on perovskite solar cells reveals ionic defect distribution}

\author{Sebastian Reichert}\thanks{These authors contributed equally to this work.}
\affiliation{Institut f\"ur Physik, Technische Universit\"at Chemnitz, 09126 Chemnitz, Germany}
\author{Jens Flemming}\thanks{These authors contributed equally to this work.}
\affiliation{Fakult\"at f\"ur Mathematik, Technische Universit\"at Chemnitz, 09126 Chemnitz, Germany}
\author{Qingzhi An}
\author{Yana Vaynzof}
\affiliation{Kirchhoff-Institut f\"ur Physik and Centre for Advanced Materials, Ruprecht-Karls-Universit\"at Heidelberg, Im Neuenheimer Feld 227, 69120 Heidelberg, Germany}
\affiliation{Technische Universität Dresden, Institut für Angewandte Physik and Centre for Advancing Electronics Dresden (cfaed), Nöthnitzer Straße 61, 01069 Dresden}
\author{Jan-Frederik Pietschmann}
\affiliation{Fakult\"at f\"ur Mathematik, Technische Universit\"at Chemnitz, 09126 Chemnitz, Germany}
\author{Carsten Deibel}\thanks{Corresponding author: deibel@physik.tu-chemnitz.de}
\affiliation{Institut f\"ur Physik, Technische Universit\"at Chemnitz, 09126 Chemnitz, Germany}

\date{\today}

\begin{abstract}
One of the key challenges for future development of efficient and stable metal halide perovskite solar cells is related to the migration of ions in these materials. Mobile ions have been linked to the observation of hysteresis in the current--voltage characteristics, shown to reduce device stability against degradation and act as recombination centers within the band gap of the active layer. In the literature one finds a broad spread of reported ionic defect parameters (e.g.\ activation energies) for seemingly similar perovskite materials, rendering the identification of the nature of these species difficult. In this work, we performed temperature dependent deep-level transient spectroscopy (DLTS) measurements on methylammonium lead iodide perovskite solar cells and developed a extended regularization algorithm for inverting the Laplace transform. Our results indicate that mobile ions form a distribution of emission rates (i.e.\ a distribution of diffusion constants) for each observed ionic species, which may be responsible for the differences in the previously reported defect parameters. Importantly, different DLTS modes such as optical and current DLTS yield the same defect distributions. Finally the comparison of our results with conventional boxcar DLTS and impedance spectroscopy (IS) verifies our evaluation algorithm.
\end{abstract}

\maketitle

\section{Introduction}

In recent years, solar cells based on organic--inorganic perovskite semiconductors have gained significant attention from the emerging photovoltaics community.\cite{Nayak2019} Perovskites exhibit many advantageous properties such as high charge carrier lifetimes, large absorption coefficients and impressive power conversion efficiencies.\cite{Leguy2016, Chouhan2017, Nrel, Jiang2017, Polman2016} However, perovskites are also rich on ionic defects, which have been shown to cause hysteresis and reduced stability.\cite{Jacobs2017,Rivkin2018, Li2016, Tessler2018} Among the various types of native point defects in methylammonium lead iodide (MAPbI\textsubscript{3}) perovskite, vacancies such as $V_\mathrm{I}^+$, $V_\mathrm{MA}^-$, and interstitials such as $\mathrm{I}_i^-$ and $\mathrm{MA}_i^+$ have been reported to be the most dominant ionic defects.\cite{Futscher2019, Meggiolaro2018, Meggiolaro2019, Yin2014, Senocrate2017, Senocrate2018} The iodine interstitial and the iodine vacancy have been shown to be the most mobile species,\cite{Senocrate2017, Azpiroz2015, Park2019} which can act as recombination centers.\cite{Meggiolaro2019, Du2014} Unfortunately, a wide variation of different ionic defect parameters such as activation energies have been reported in the literature\cite{Futscher2019,Eames2015, Azpiroz2015,Yang2016,Yin2014} even for the same types of perovskite solar cells. Therefore, the identification of the nature of these species is difficult. This is particularly noticeable for iodine migration. Theoretical calculations predicted an activation energy of $~0.1~\mathrm{eV}$ for iodine migration.\cite{Azpiroz2015} \textcite{Duan2015} found an activation energy of $0.16~\mathrm{eV}$ by admittance spectroscopy. A slightly higher activation energy was determined by transient ionic-drift measurements to $0.29~\mathrm{eV}$.\cite{Futscher2019} Chronophotoampereometry measurements revealed activation energies of $0.6~\mathrm{eV}$.\cite{Eames2015} Overall, the results reported in literature for the activation energy of iodine ions varies within $0.5~\mathrm{eV}$.

In this work, we performed deep-level transient spectroscopy (DLTS) measurements to examine the properties of mobile ions in methylammonium lead iodide solar cells. The capacitance transients obtained by DLTS typically represent the response of a wide range of emission rates. The analysis of such multi-exponential processes is challenging due to the commonly unknown number of single components. It is further complicated by superposition of closely adjacent time constants in the presence of noise. Herein, we introduce an enhanced regularization algorithm for the inverse Laplace transform which significantly simplifies the evaluation of DLTS measurements. The application of this algorithm to temperature dependent DLTS data reveals three distinct ionic species, each exhibiting a distribution of diffusion coefficients. A comparison with conventional boxcar evaluation verifies the results obtained by the regularization algorithm. Moreover, the results are consistent when different DLTS modes such as reverse, optical and current DLTS are applied to obtain a more detailed understanding of the distributions of diffusion coefficients. Our measurements indicate two anion and one cation distributions, which we assign to $\mathrm{I}_i^-$, $V_\mathrm{MA}^-$ and $\mathrm{MA}_i^+$,  respectively. The diffusion coefficients of $\mathrm{I}_i^-$ and $V_\mathrm{MA}^-$ were found to be three orders of magnitude higher than that of the $\mathrm{MA}_i^+$ species, indicating the high mobility of $\mathrm{I}_i^-$ and $V_\mathrm{MA}^-$ ions. The ion concentrations of all species are within the same order of magnitude, so that all species have a comparably high impact on the device properties. It is noteworthy that the observed ionic defect distributions can explain the large deviations of ionic defect parameters previously reported in literature. Finally, unraveling these defect distributions can help to understand how to avoid their formation during the active layer fabrication, in order to improve the performance and stability of perovskite solar cells. Alternatively, it may assist with the development of passivation strategies tailored to certain types of defects.

%%%%%%%%%%%%%%%%%%%%%%%%%%%%%%%%%%%%%%%%%%%%%%%%%%%Experimental

\section{Methods and Theory}\label{sec:methods}

\subsection{Sample preparation and current density--voltage characteristics}

Pre-patterned indium tin oxide (ITO) coated glass substrates (PsiOTech Ltd., $15~\Omega/\mathrm{sqr}$) were ultrasonically cleaned with $2~\%$ Hellmanex detergent, deionized water, acetone, and isopropanol, followed by 10~min oxygen plasma treatment. Modified poly(3,4-ethylene-dioxythiophene):poly(styrenesulfonate) (m-PEDOT:PSS) was spin cast on the clean substrates at 4000~rpm for 30~s and annealed at $150~^\circ\mathrm{C}$ for 15~min to act as hole transport layer.\cite{Zuo2016} The MAPbI\textsubscript{3} active layer was formed using the lead acetate trihydrate route following previous works.\cite{An2019,Fassl2018} In short, the perovskite solution (at a stoichiometry of 1:3.00 MAI:PbAc\textsubscript{2}) was spin cast at 2000~rpm for 60~s in a dry air filled glovebox (relative humidity $<0.5~\%$). After blowing 25~s and drying 5~min, the as-spun films were annealed at $100~^\circ\mathrm{C}$ for 5~min forming a uniform perovskite layer. The prepared samples were transferred to a nitrogen filled glove box, where an electron transport layer [6,6]-phenyl-C61-butyric acid methylester (PC\textsubscript{60}BM), 20~mg/ml dissolved in chlorobenzene, was dynamically spin cast at 2000~rpm for 30~s on the perovskite layer followed by a 10~min annealing at $100~^\circ\mathrm{C}$. Sequentially, a bathocuproine (BCP), 0.5~mg/ml dissolved in isopropanol, hole blocking layer was spin cast on top of the PC\textsubscript{60}BM. The device was completed with a thermally evaporated 80~nm thick silver layer.

The current density--voltage (jV) characteristics were measured by a computer controlled Keithley 2450 Source Measure Unit under simulated AM 1.5 sunlight with $100~\mathrm{mW/cm}^2$ irradiation (Abet Sun 3000 Class AAA solar simulator). The light intensity was calibrated with a Si reference cell (NIST traceable, VLSI) and corrected by measuring the spectral mismatch between the solar spectrum, the spectral response of the perovskite solar cell and the reference cell.

\subsection{Ionic defects in perovskite solar cells}

Ionic defects have a major impact on device stability and performance, and can lead to deep-level trapping of charge carriers.\cite{Fassl2018, Du2014, Meggiolaro2019, Zhao2017, Kim2017, Cheng2016} It is therefore necessary to include the ionic responses when analyzing the solar cell behavior. When describing the capacitance of a perovskite solar cell, the capacitive response caused by ions ($C_\mathrm{ion}$) has to be added to the geometrical capacitance ($C_\mathrm{0}$, considering the solar cells as a classical parallel plate capacitors in dependence of the permittivity $\epsilon_\mathrm{R}$, the active area $A$ and thickness $L$ of the solar cell) and the capacitance caused by charge carriers trapped in defect states ($C_\mathrm{trap}$):
\begin{equation}
    \label{eq:C_tot}
    C_\mathrm{tot}=C_\mathrm{0}+C_\mathrm{trap}+C_\mathrm{ion}.
\end{equation}

As recent studies show, mobile ions are a dominant process for limiting device performance and stability. Classical semiconductor deep-level defects are less significant since the low concentration of such defects compared with the high concentration of mobile ionic species.\cite{Futscher2019, Cheng2016, Song2017, Sherkar2017} These results are confirmed by theoretical calculations and simulations showing that the deep-level trapping of charge carriers is quite unlikely compared to ionic movement.\cite{Yin2014,Senocrate2018,Buin2015, Rakita2019} Mobile ions must overcome an activation energy $E_\mathrm{A}$ in order to hop between localized states within the perovskite lattice.\cite{Shao2016, Ferdani2019, Frost2016} A temperature dependent emission rate $e_\mathrm{t}$ can be defined for describing this hopping process:\cite{Heiser1993, Zamouche1995, Futscher2019}
\begin{equation} 
    \label{eq:en_D}
    e_\mathrm{t}=\frac{e^2DN_\mathrm{eff}}{k_\mathrm{B}T\epsilon_\mathrm{0}\epsilon_\mathrm{R}},
\end{equation}
where $e$ is the elementary charge, $D$ the ion-diffusion coefficient, $N_\mathrm{eff}$ the effective doping density, $k_\mathrm{B}$ the Boltzmann constant, $T$ the temperature and $\epsilon_\mathrm{0}$ the absolute dielectric permittivity. The ion diffusion coefficient is linked to the activation energy $E_\mathrm{A}$ by
\begin{equation} 
    \label{eq:D}
    D=D_\mathrm{0}\exp{\left(-\frac{E_\mathrm{A}}{k_\mathrm{B}T}\right)},
\end{equation}
with the diffusion coefficient at infinite temperatures $D_\mathrm{0}$.

\subsection{Characterization of mobile ions with DLTS}

\begin{figure}[h]
    \includegraphics*[scale=0.6]{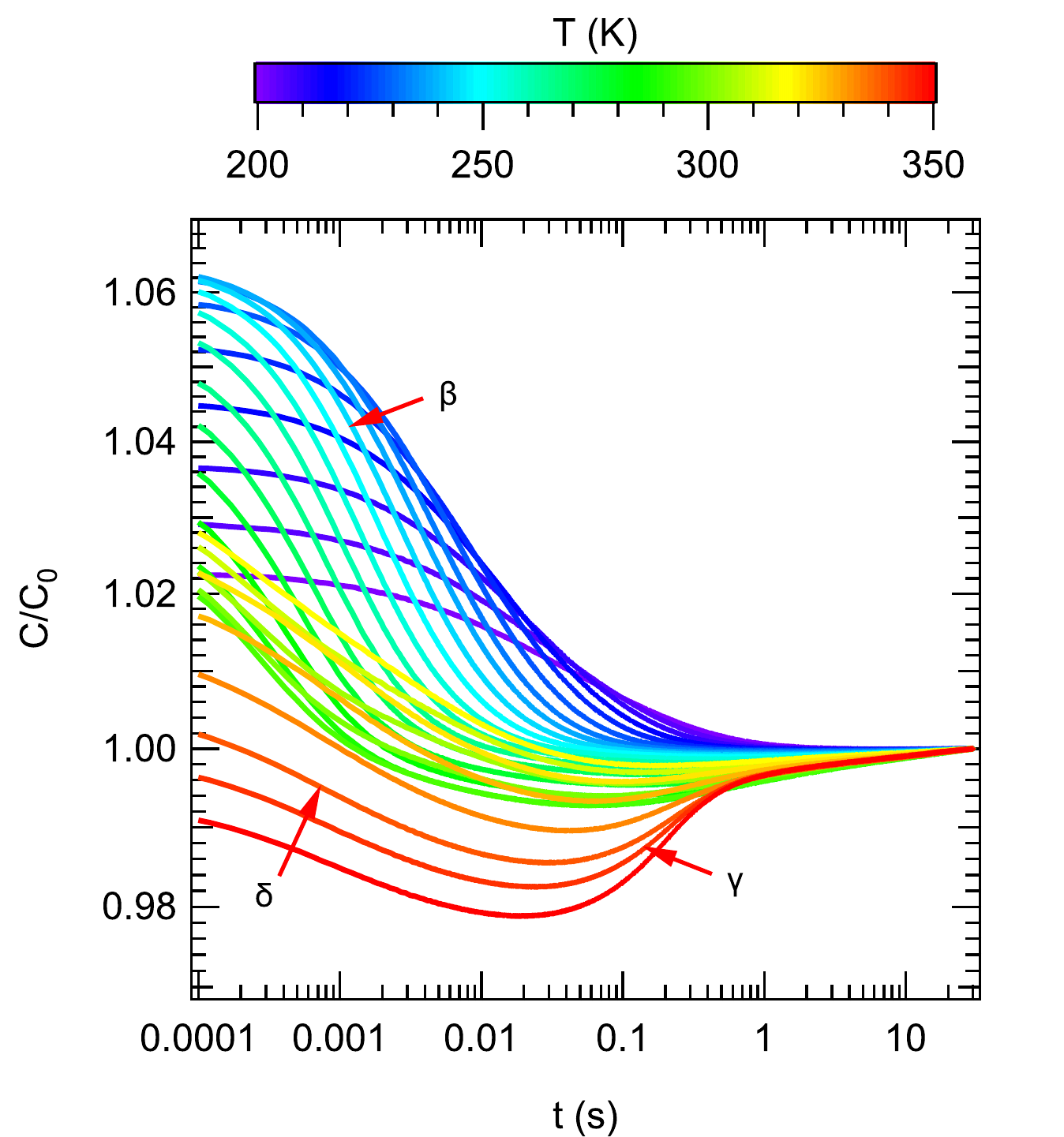}
    \caption{DLTS measurements on MAPbI\textsubscript{3} solar cell for different temperatures from 200~K to 350~K in 5~K steps. The capacitance transients were obtained after pulsing the sample from 0~V to 1~V for 100~ms.}
    \label{fig:Ct}
\end{figure}

An effective technique to obtain specific information about mobile ions in perovskite solar cells is DLTS.\cite{Lang1974} This technique was originally conceived for measuring electronic deep-level defects, but mobile ions can also be investigated due to their dependence on external voltages and capacitive response. The device under test is initially biased at zero volt. During a voltage or light pulse, mobile ions previously located at an interface between perovskite layer and transport layer are pushed into the perovskite until they reach a new steady state. After the pulse is finished, mobile ions relax back to the steady state condition by returning to the interfaces which gives rise to a capacitance transient
\begin{equation} 
    \label{eq:Ct}
    C_\mathrm{t}=C_{\infty}\pm \Delta C \exp{(-e_\mathrm{t}t)},
\end{equation}
where $C_{\infty}$ is the steady state capacitance. The ion concentration $N_\mathrm{ion}$ is proportional to $\Delta C$,
\begin{equation} 
    \label{eq:DC}
    \Delta C\propto \frac{N_\mathrm{ion}}{N_\mathrm{eff}}C_{\infty}\quad \textrm{  if  } N_\mathrm{ion}\ll N_\mathrm{eff}.
\end{equation}
The DLTS setup used in this work consists of a Zurich Instruments MFLI lock-in amplifier with MF-IA and MF-MD options, a Keysight Technologies 33600A function generator and a cryo probe station Janis ST500 with a Lakeshore 336 temperature controller. The capacitance transients (Fig.~\ref{fig:Ct}) were recorded in the temperature range of 200~K to 350~K in 5~K steps, controlled accurately within 0.01~K, using liquid nitrogen for cooling. DLTS measurements were performed using an AC frequency of 80~kHz with amplitude of $V_\mathrm{ac}=20~\mathrm{mV}$ and biasing the perovskite device from 0~V to 1~V for 100~ms. The transients were measured over 30~s and averaged over 35 single measurements.

For the extraction of characteristic information from DLTS data several methods are available. Most commonly, DLTS is evaluated using the boxcar method.\cite{Lang1974, Khan2015, Rau2016} Plotting the extracted emission rates on logarithmic scale over 1000/$T$ in an Arrhenius diagram results in straight lines according to Eqn.~(\ref{eq:en}). These straight lines contain information about the activation energy for ion migration (slope) and the diffusion coefficient (y-intercept). The boxcar evaluation is a valuable technique to obtain emission rates without presuming any specific model. The disadvantage of the boxcar evaluation is the low time resolution especially in presence of noise, which makes it difficult to distinguish between two closely spaced rates (see Supplemental Material, part I). A more sensitive evaluation method is achieved by the inverse Laplace transform. The exponential capacitance transient $C(t)$ can be considered as the Laplace transform of a function in the transform space $e_\mathrm{t}$.\cite{Tin2002,Schroder2005} That means $C(t)$ is the Laplace transform of $g(e_\mathrm{t})$ and can be expressed as
\begin{equation} 
    \label{eq:LT1}
    C(t)=\sum_i C_i\exp{(-e_{\mathrm{t}_i}t)}=\int_{0}^{\infty} g(e_\mathrm{t})\exp{(-e_\mathrm{t}t)} \mathrm{d}e_\mathrm{t}.
\end{equation}
The inverse Laplace transform of $C(t)$ is subsequently given by the spectral function $g(e_\mathrm{t})$
\begin{equation} 
    \label{eq:LT2}
    g(e_\mathrm{t})=\sum_i C_i\delta(e_\mathrm{t}-e_{\mathrm{t}_i}).
\end{equation}
Peaks at $e_\mathrm{t}=e_{\mathrm{t}_i}$ can be obtained by plotting $g(e_\mathrm{t})$ versus $e_\mathrm{t}$ and by transferring the emission rates into an Arrhenius diagram according to Eqn.~(\ref{eq:en}).

Since inverting the Laplace transform is an ill-posed problem in the sense that small errors in measurements may yield large errors in the inversion, regularization of the inversion process is necessary. A first numerical implementation was developed by \textcite{Provencher1982} (Contin) and later modified (Ftikreg).\cite{Provencher1982,Weese1992,Weese1993,Honerkamp1990,Dobaczewski2004} For successfully using these Tikhonov regularization-based methods, a high signal-to-noise (SNR) ratio in the range of 1000 is required.\cite{Dobaczewski2004} A large number of transients have to be averaged to achieve such a high SNR. %known from experimental experiences
This issue is challenging because of the long transient time of about 30 seconds necessary for measuring ion movement in perovskite solar cells. The long recording time in combination with averaging over a high member of single measurements for each transient would result in long measurement times for each temperature step. To overcome this issue, we implemented an improved algorithm to regularize the inverse Laplace transform, implemented in the modern programming language Python. The enhanced method is fast and easy in use due to the lower amount of parameters necessary to achieve reliable results.

\subsection{Numerical inversion of Laplace transform via Regularized Sparse Laplace Spectrum RegSLapS}\label{RegSLapS}

Inverting the Laplace transform in Eqn.~(\ref{eq:LT1}) is an ill-posed problem in the sense that very small deviations in the data $C$ (noise, round-off errors) may cause extremely large errors in the solution $g$. Equivalently, very different spectral functions $g$ may have almost the same Laplace transform and, thus, cannot be reconstructed solely from data $C$ as the very small difference of their Laplace transforms cannot be distinguished from noise.

To obtain reasonable reconstruction results for ill-posed problems additional information about expected solutions has to be taken into account. One standard approach is Tikhonov's regularization method:
\begin{equation}\label{eq:tikh}
\|L\,g-C\|^2+\alpha\,\|g\|^2\to\min_g.
\end{equation}
Here $\|\cdot\|$ denotes the mean-square norm of a function and $L$ is the Laplace transform, a linear mapping.
The regularization parameter $\alpha>0$ controls the trade-off between data fitting and regularization, where the latter is a consequence of the penalty term $\|g\|^2$. See Engl et al.\cite{EngHanNeu96} for details on Tikhonov regularization.

Additional information added by this standard Tikhonov approach is that the sought-for spectral function has a small norm. The Contin algorithm for processing DLTS measurements uses the squared norm of the second derivative $g''$ as penalty instead of $\|g\|^2$. This leads to very smooth (low curvature) spectral functions $g$. The advantage of the Contin approach is that its numerical realization is rather simple, since minimizing the corresponding Tikhonov-like functional is achieved by simply solving a well-posed system of linear equations.

The approach presented here differs significantly from standard Tikhonov and Contin. Since in addition to smooth sections, Laplace spectra may also contain sharp peaks, it is undesirable to apply methods that yield only very smooth results. Deciding whether a smooth regularized spectrum originates from sharp peaks or from smooth distributions is almost impossible. But as will be shown in Sec.~\ref{sec:results} from a regularized spectrum containing peaks, it can be easily determined if the real spectrum was smooth or contained peaks.

Functions consisting of a small number of peaks can be mathematically encoded as preferring a function which is zero on large parts of its domain. Such functions are the usual outcome of a method called $\ell^1$-regularization, see \textcite{DauDefDem04} This method involves solving the minimization problem
\begin{equation*}
    \|L\,g-C\|^2+\alpha\,\int_0^\infty |g(e_\mathrm{t})|\,\mathrm{d}e_\mathrm{t}\to\min_g.
\end{equation*}
The problem is again of Tikhonov-type, but is accompanied with a non-differentiable penalty due to the absolute value inside the integral. This non-differentiability makes numerical minimization more challenging.

Our implementation RegSLapS is based on fast iterative soft-threshold algorithm (FISTA),\cite{BecTeb08} which is essentially a projected gradient method. To describe the algorithm we replace the spectral function $g$ by a finite vector of function values at fixed grid points. Then the Laplace transform $L$ becomes a matrix which maps $g$ to a finite data vector $C$. We use the following algorithm
\begin{itemize}
    \item initialization: fix $\alpha>0$ (details below), set $g$ to zero everywhere, set $g_{\mathrm{old}}:=g$
	\item do until convergence:
	\begin{itemize}
		\item mixing step: \quad$g_{\mathrm{mix}}:=g+\frac{i-1}{i+2}\,(g-g_{\mathrm{old}})$\\
	(where $i$ is the iteration counter)
		\item calculate gradient of fitting functional:\\
	$d:=L^{\mathrm{T}}\,(L\,g_{\mathrm{mix}}-C)$
		\item calculate step size: \quad$s:=\frac{\|d\|^2}{\|L\,d\|^2}$
		\item gradient step: \quad$\tilde{g}:=g_{\mathrm{mix}}-s\,d$
		\item projection step: for all $k$ set\\
			$g_{\mathrm{proj}}:=(\mathrm{sign}\, \tilde{g}_k)\,\max\{0,\,|\tilde{g}_k|-s\,\alpha\}$
		\item update: set \quad$g_{\mathrm{old}}:=g$\quad and \quad$g:=g_{\mathrm{proj}}$
	\end{itemize}
\end{itemize}

The choice of the regularization parameter $\alpha$ is crucial and difficult. However, DLTS data is very smooth for prolonged times $t$, which allows to obtain a reasonable estimate for the data's noise level as the difference between measured data and a smoothed version of the measured data. Based on this noise level we can apply Morozow's discrepancy principle\cite{Mor66} as follows:
\begin{itemize}
		\item choose some large initial $\alpha$
		\item calculate regularized solution, denoted by $g_\alpha$
		\item do until $\|L\,g_\alpha-C\|\leq k\eta$ (where $\eta$ is the noise level and $k>1$ a constant close to one):
		\begin{itemize}
			\item decrease $\alpha$ by some fixed factor
			\item calculate regularized solution $g_\alpha$
		\end{itemize}
		\item last $g_\alpha$ is the result of the method
	\end{itemize}

The estimate of the spectral function $g$ allows for peaks to be simply identified as the grid points for which $g$ is not zero. Tests using synthetic data confirm that the area under the curve is very close to the height of a $\delta$-peak in the exact spectral function. For a verification of our algorithm, we tried to reproduce pre-factor and time constants of several synthetic multi exponential decays (see Supplemental Material, part I). In all cases, RegSLapS was able to reproduce the parameters of the synthetic data accurately. This approach serves as a test of validity of RegSLapS.

%%%%%%%%%%%%%%%%%%%%%%%%%%%%%%%%%%%%%%%%%%%%%%%%%%%Results

\section{Results and discussion}\label{sec:results}

The photovoltaic performance of the devices investigated here is that of typical MAPbI\textsubscript{3} solar cells. The current density–voltage characteristics are shown in Supplemental Material, part II. The device exhibits a current density of $21.6~\mathrm{mA/cm}^2$, open circuit voltage of $0.96~\mathrm{V}$, fill factor of $76.5~\%$ and a power conversion efficiency of $15.8~\%$. As reported in literature, jV measurements are often affected by hysteresis.\cite{Snaith2014, Tress2015, Unger2014} The origin of hysteresis is attributed to mobile ion within the active layer of perovskite solar cells.\cite{Contreras2016, Weber2018, Lee2019, Kim2018} The jV data show almost no hysteresis, but corresponding to the results of \textcite{Calado2016} ion migration can also take place in devices with minimal hysteresis.

In Supplemental Material part III, the rate window evaluation of the DLTS transients (Fig.~\ref{fig:Ct}) is shown. Three ion-related peaks are visible, which are labeled as $\beta$, $\gamma$ and $\delta$. The sign of the peaks are related to the charge state of the contributing ions. Defects $\beta$ and $\delta$ are anions whereas $\gamma$ is a cation. We determine the two low-frequency responses below 10~Hz to correspond to the same defect $\gamma$. We cannot exclude that the defects $\beta$ and $\delta$ correspond to the same ionic origin. However, as a working hypothesis, we assume that they are different ionic defects as we did not observe a change in the activation energy or diffusion coefficient according to \textcite{Futscher2019} in the IS data at the tetragonal-cubic phase-transition temperature at 327~K.\cite{Saidi2016}

\begin{figure}[tb]
    \includegraphics*[scale=0.6]{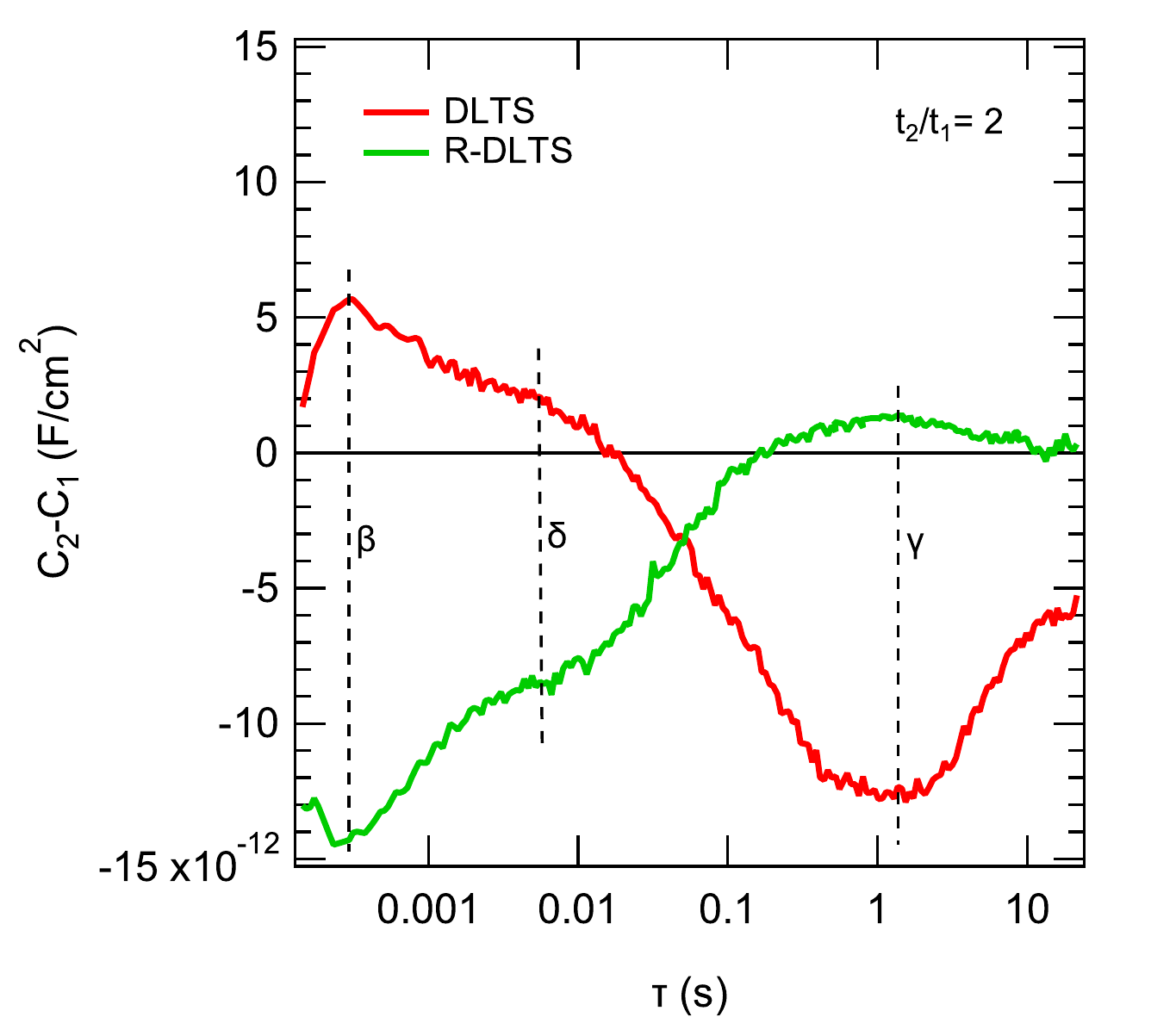}
    \caption{Conventional DLTS and reverse DLTS evaluation via boxcar at room temperature with a rate window of $t_2/t_1=2$. The emission rates are equal for all defects which indicates the underlying ionic origin.}
    \label{fig:RDLTS}
\end{figure}
We distinguished the nature of defects, electronic or ionic, by applying reverse DLTS.\cite{Li1983, Li1985} In this technique, the filling pulse is reversed, meaning that the voltage pulse goes from 1~V to 0~V. Other parameters such as AC frequency and filling pulse parameters remain constant. As a consequence for electronic traps, the capture process of charge carriers in trap states is measured in contrast to conventional DLTS measurements. Since the emission rate is much lower than the capture rate, the latter (determined by reverse DLTS) is expected to be higher than the emission rate (obtained by conventional DLTS) if deep-level trap states are being measured. The situation is different, when mobile ions respond to the voltage pulse. When no external voltage is applied, mobile ions are at equilibrium at the interfaces of the perovskite layer due to the internal field.\cite{Meggiolaro2019, Futscher2019} As a result, the mobile ions will be pushed into the perovskite layer by the voltage pulse. For ions it can be expected that this process is approximately as fast as the back drift of the ions to the interfaces. We measured reverse DLTS at room temperature and plotted the boxcar evaluation for a rate window of $t_2/t_1=2$ in Fig.~\ref{fig:RDLTS}. Since the emission rates for all defects are comparable for conventional and reverse DLTS, we can conclude that the measured emission rates originate from mobile ions and not charge carriers such as electron and holes.

A defect signature can be obtained by measuring the emission rate $e_\mathrm{t}(T)$ at different temperatures, as the emission rate for ions is thermally activated according to Eqn.~(\ref{eq:en_D}) and (\ref{eq:D}):\cite{Futscher2019, Azpiroz2015, Yang2016}
\begin{equation} 
    \label{eq:en}
    e_\mathrm{t}=\frac{e^2N_\mathrm{eff}D_\mathrm{0}}{k_\mathrm{B}T\epsilon_\mathrm{0}\epsilon_\mathrm{R}}\exp{\left(-\frac{E_\mathrm{A}}{k_\mathrm{B}T}\right)}.
\end{equation}
Emission rates for each temperature were plotted in Fig.~\ref{fig:arrhenius_C}, obtained by evaluating the peaks visible in the rate window plot (see Supplemental Material, part III). Eqn.~(\ref{eq:en}) was used to fit the data, in order to determine the activation energy $E_\mathrm{A}$ and diffusion coefficient $D$ with Eqn.~(\ref{eq:D}). For calculating these defect parameters the effective doping density and permittivity were determined by capacitance--voltage measurements (see Supplemental Material, part IV, the solar cells was pre-biased at 1~V for 60~s and rapidly swept with 3~V/s in reverse direction\cite{Fischer2018}). The results are summarized in Table~\ref{tab:table1}.
\begin{table}
	\caption{\label{tab:table1} Activation energy $E_\text{A}$, diffusion coefficient $D_\mathrm{300K}$ at 300~K and ion concentration $N_\text{ion}$ obtained by boxcar evaluation.}
	\begin{ruledtabular}
	\begin{tabular}{ccccc}
	defect & ion & $E_\text{A}~\mathrm{(eV)}$ & $D_\mathrm{300K}~\mathrm{(cm^2s^{-1})}$ & $N_\text{ion}~\mathrm{(cm^{-3})}$ \\
	\hline
	$\beta$ & $V_\text{MA}^-$ & 0.37 & $2\cdot10^{-7}$ & $6\cdot10^{14}$ \\
	$\gamma$ & $\text{MA}_i^+$ & 0.37 & $1\cdot10^{-11}$ & $2\cdot10^{14}$ \\
	$\delta$ & $\text{I}_i^-$ & 0.19 & $5\cdot10^{-9}$ & $9\cdot10^{13}$ \\
	\end{tabular}
	\end{ruledtabular}
\end{table}
We determined the activation energy of $\beta$ to 0.37~eV, for $\gamma$ to 0.37~eV and for $\delta$ to 0.19~eV. These results are in agreement with the results of \textcite{Park2019} and \textcite{Yin2014} which supports our findings that the measured ionic defects have low activation energies. The diffusion coefficient at 300~K for $\delta$ was determined to be $5\cdot10^{-9}~\mathrm{cm^2s^{-1}}$. For $\beta$ the diffusion coefficient at 300~K is $2\cdot10^{-7}~\mathrm{cm^2s^{-1}}$, which is several orders of magnitude higher when compared to that of $\gamma$ with $1\cdot10^{-11}~\mathrm{cm^2s^{-1}}$. The determined diffusion coefficients at 300~K are in the range of $10^{-7}$--$10^{-12}~\mathrm{cm^2s^{-1}}$, consistent with values reported in literature.\cite{Yang2015, Eames2015, Senocrate2017, Futscher2019, Musiienko2019} Ion concentrations were found to be $6\cdot10^{14}~\mathrm{cm^{-3}}$ for $\beta$, $2\cdot10^{14}~\mathrm{cm^{-3}}$ for $\gamma$, and $9\cdot10^{13}~\mathrm{cm^{-3}}$ for $\delta$.

By comparison with literature we made a conceivable scenario for identifying the measured defects with mobile species in perovskite solar cells. First we compared the results of our measurements with those reported by \textcite{Futscher2019} (see Supplemental Material, part V). This allowed us to attribute defect $\gamma$ to $\text{MA}_i^+$ since both the position and the sign of both measurements are in good agreement. However, in contrast to the interpretation of phase transition induced changes of the diffusion coefficient and activation energy, our results predict that different parts of the same defect distribution were measured. This is in agreement with the two measured $\text{MA}_i^+$ responses determined by \textcite{Futscher2019} (see Supplemental Material, part V, C1 and C2) which belong to the same defect distribution. The identification of $\beta$ and $\delta$ is less certain. The defect attributed to $\text{I}_i^-$ (see A1 in Supplemental Material, part V) measured by \textcite{Futscher2019} is indeed close to the edge of distribution $\beta$. However, both distribution $\beta$ and $\delta$ are very close to one another and it cannot be excluded that the distribution $\delta$ is also present at low temperatures and frequencies. Consequently, the identification of $\beta$ and $\delta$ is not possible by comparing with the results of \textcite{Futscher2019}. Instead, we compared our results with the DFT calculations of \textcite{Yang2016} The reported activation energy of $\text{MA}_i^+$ ions (ab-plane: $0.38~\mathrm{eV}$, c-axis: $0.48~\mathrm{eV}$) is in good agreement with our findings. Furthermore, the reported diffusion barrier of \textcite{Yang2016} for $\text{I}_i^-$ ions (ab-plane: $0.19~\mathrm{eV}$, c-axis: $0.33~\mathrm{eV}$) are in the same range as our result for defect $\delta$. Therefore we attributed $\delta$ to $\text{I}_i^-$. Since the only negative ion among all likely native point defects left is $V_\text{MA}^-$, we assign $\beta$ to $V_\text{MA}^-$. Noticeably, the activation energy of \textcite{Yang2016} for $V_\text{MA}^-$ (ab-plane: $0.62~\mathrm{eV}$, c-axis: $0.89~\mathrm{eV}$) does not agree with our findings and at the moment we have not yet an explanation for that discrepancy. Moreover we would expect comparable diffusion coefficients for both $\text{MA}_i^+$ and $V_\text{MA}^-$ in contrast to our observation (Tab.~\ref{tab:table1}). However, our assignment is in good agreements with the high mobility reported for anions which is significantly higher than the mobility of cations.\cite{Meggiolaro2018,Park2019, Azpiroz2015,Senocrate2017, Senocrate2018} 

\begin{figure*}[t]
    \includegraphics*[scale=0.7]{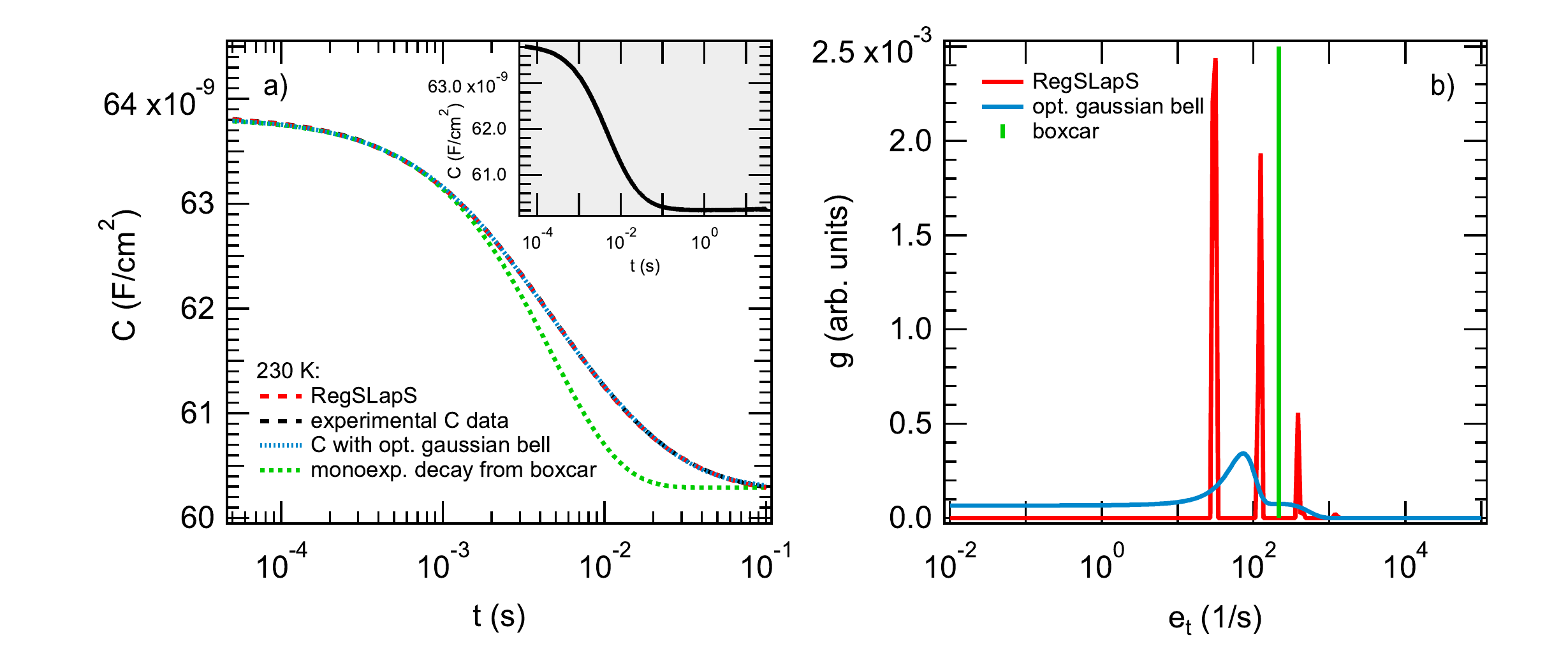}
    \caption{(a) Capacitance transients of experimental data, reconstructed with the emission rates calculated with RegSLapS and reconstructed by fitting Gaussian defect distributions. The data were cut at 0.1~s to avoid the influence of $\gamma$ on fitting Gaussian distributions to the emission rate spectrum, which is still present even for low temperatures (inset). (b) Calculated emission rate spectrum of RegSLapS with fit of Gaussian distributions.}
    \label{fig:distribution}
\end{figure*}
It is important to note that the emission rates obtained by boxcar evaluation for $\beta$, $\gamma$ and $\delta$ cannot describe the capacitance transients completely. As shown in Fig.~\ref{fig:distribution}a, we highlighted this observation using the 230~K capacitance transient up to 0.1~s exemplary for $\beta$.  We chose this particular transient as a compromise between good signal-to-noise ratio and absence of the response of the other defects $\gamma$ and $\delta$. The mono-exponential decay with the emission rate obtained by boxcar evaluation is insufficient to fit the experimental data. The ambiguity between experimental result and the capacitance transient reconstructed very likely indicates a defect distribution, as a distribution of emission rates is necessary to describe the experimental transient. Consequently, by fitting a defect distribution consisting of two Gaussian distributions to the capacitance transient, which yield the emission rate spectrum Fig.~\ref{fig:distribution}b, a good agreement with the experimental capacitance transient can be obtained. The presence of a defect distribution would offer a reasonable explanation for the different and partly inconsistent defect parameters reported for perovskite solar cells in literature.\cite{Meggiolaro2019, Eames2015, Azpiroz2015, Futscher2019, Yang2017} Different experimental methods and conditions could lead to probing different  parts of the defect distribution. We propose that the broad distribution of ionic defects might originate from the non-constant internal electric field in perovskite solar cells, in part caused by ion migration.

To exploit the higher resolution offered by inverse Laplace transform, we used our RegSLapS algorithm as described in Sec.~\ref{RegSLapS}. At first we considered again the example shown in Fig.~\ref{fig:distribution}. With RegSLapS we obtained three different sharp peaks, which are roughly equidistant spaced (see Fig.~\ref{fig:distribution}b). As shown in Fig.~\ref{fig:distribution}a, the three sharp emission rates given by RegSLapS as well as the distribution of emission rates were able to fit the data adequately. That means, we cannot distinguish, whether the defect distribution or the emission rates given by RegSLapS are more suitable the explain the experimental result. The reason is that the inverse Laplace transform is an ill-posed problem. However, we tested RegSLapS with a synthetic Gaussian distribution and obtained equidistant peaks instead of the defect distribution in the resultant spectra (see Supplemental Material, part I). This indicates that the measured emission rates with equidistant spacing correspond to a defect distribution. To achieve short calculation times, the algorithm described in Sec.~\ref{sec:methods} was adjusted to keep the amount of calculated solutions low by preferring zeros in the solution space. Therefore, the calculated emission rate spectrum by RegSLapS results in $e_\mathrm{t}$ with equidistant spacing instead of a continuous distribution of emission rates. Moreover, recently it has been reported that perovskite crystal grain orientation influences the rate of lateral ionic transport in spin cast MAPbI\textsubscript{3} films.\cite{Fassl2018_2} In this work, photoluminescence microscopy was utilized to track the migration of ions upon the application of an electric field, with the results showing a broad quenching front being present only for films with a random grain orientation. Modelling of these results was only possible using a large distribution of diffusion coefficients for the same ionic defect. It is thus possible, that different grain orientations in the vertical solar cell devices also result in a wide distribution of ionic transport rates, in agreement with the results presented here. In summary, based on first the fact that our transient data cannot be reproduced with mono exponential decays, second that RegSLapS will yield distinct, equal spaced rates when a distribution of rates is present, and third the information from \textcite{Fassl2018_2} in which the data could only be explained by a distribution of diffusion coefficients, we expect that all of our observed ion migration rates belong to broad distributions.

With this finding, we evaluated the capacitance transients from Fig.~\ref{fig:Ct} and plotted the resulting emission rates in Supplemental Material, part VI. As mentioned above, inverse Laplace transform is an ill-posed problem, so that artifacts can occur within the spectrum calculated by RegSLapS. This means that only calculated emission rates are physically justifiable, which follow the emission rates equation Eqn.~(\ref{eq:en}).
\begin{figure}[h]
    \includegraphics*[scale=0.60]{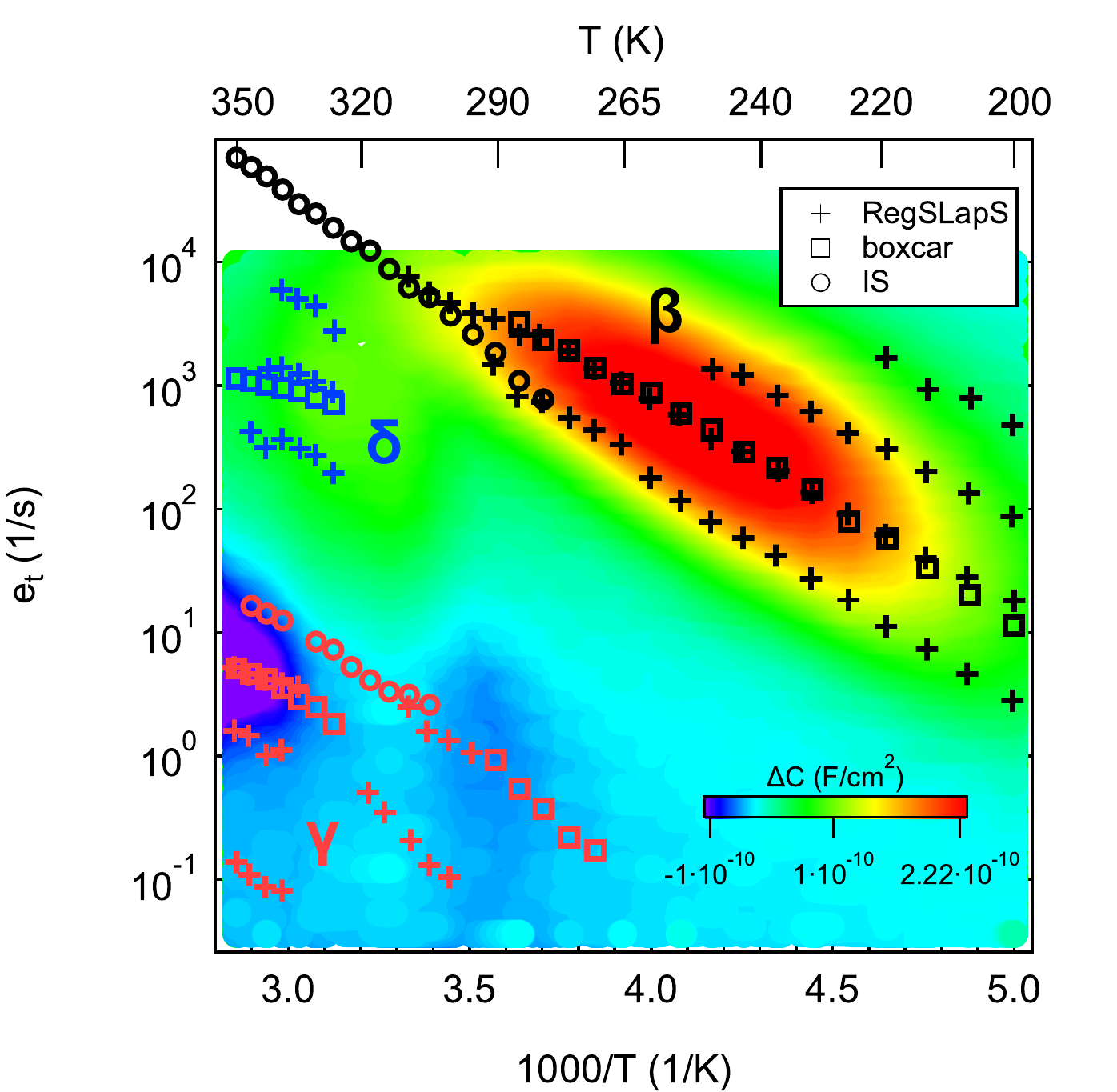}
    \caption{Comparison of emission rates obtained by conventional boxcar evaluation (contour plot of part III in Supplemental Material), by inverse Laplace transform via RegSLapS and by impedance spectroscopy (IS). Visible are three different ionic defects $\beta$, $\gamma$ and $\delta$. Parallel straight lines are related to ionic distributions.}
    \label{fig:arrhenius_C}
\end{figure}
Emission rates that satisfy this criterion  are shown in Fig.~\ref{fig:arrhenius_C} resulting in a set of straight lines for each defect. As discussed above the emission rates that differ in the vertical offset are linked to a defect distribution.

The contour plot of the boxcar data (see Supplemental Material, part III) in Fig.~\ref{fig:arrhenius_C} shows an excellent agreement with the equidistantly spaced emission rates, since all emission rates determined using RegSLapS are within the bounds of the broad peak obtained by the boxcar evaluation. Only the maxima of the emission rate peaks can be extracted by boxcar, likely close to the peak of the experimental emission rate distribution. The agreement between our algorithm RegSLapS and the established but less sensitive boxcar evaluation confirms the validity of RegSLapS as well.

To verify our results, we performed impedance spectroscopy (IS) measurements (see Supplemental Material, part VII), which were analyzed and added to the contour plot (Fig~\ref{fig:arrhenius_C}). As IS is a small-signal technique, the perovskite solar cells are closer to equilibrium conditions during the measurement compared with DLTS, where voltage pulses have a strong effect on the internal electric fields which influence the ion migration. The comparison of IS with DLTS is therefore very helpful to verify the results obtained by DLTS. As shown in Fig.~\ref{fig:arrhenius_C} there is a good agreement of the emission rates of defect $\beta$, whereas defect $\delta$ is not visible. In the case of defect $\gamma$ only one part of the defect distribution can be measured, in contrast to DLTS, because of the lower sensitivity of IS for low emission rates.

Additionally to capacitance DLTS measurements (Fig.~\ref{fig:Ct}), we used different DLTS modes such as optical DLTS (O-DLTS, see Supplemental Material, part VIII) and current DLTS (I-DLTS, see Supplemental Material, part IX) to verify our previous results and to learn more about the nature of these ionic defects. For O-DLTS the solar cell is probed using a light pulse ($250~\mathrm{mW/cm}^2$ for 100~ms) instead of a voltage pulse. In the case of I-DLTS measurements the current response of the solar cells after applying the voltage pulse was measured simultaneously with the capacitance transient. Parameters such as filling pulse duration, transient measurement time and the number of averages corresponds to that of capacitance DLTS. As shown in Supplemental Material part X, all DLTS modes gives access to the same defect distributions, whereas with each DLTS mode another part of the distribution is measurable. In the case of our I-DLTS measurements, more than one component of the defect distribution of $\gamma$ and $\delta$ is visible even in the boxcar evaluation due to the higher resolution of I-DLTS measurements for higher emission rates in comparison to C-DLTS measurements. This verifies our result on the ionic defect distributions using RegSLapS. The existence of transients in O-DLTS measurements indicate the ion movements caused by light interaction, since light induces free charge carriers with impact on the internal electrical field.\cite{Eames2015} The result indicates the implication of mobile ions in optoelectronic measurements. Since mobile ions have an impact on the internal electrical field, they may influence the charge carrier extraction and recombination in the perovskite solar cell. Moreover, additional ionic response can occur which can be mistakenly interpreted as extraction or recombination rate. This must be taken into account for optoelectronic measurements such as IMPS/IMVS, photocurrent decay and photovoltage decay measurements.\cite{Eames2015, Contreras2016, Guilln2014, Pockett2017, Ebadi2019, Calado2016, Ravishankar2019} 

%%%%%%%%%%%%%%%%%%%%%%%%%%%%%%%%%%%%%%%%%%%%%%%%%%%Conclusions

\section{Summary}

In this work we performed DLTS measurements on MAPbI\textsubscript{3} perovskite solar cells and analyzed them using a extended algorithm for inverse Laplace transform. Our results indicate the presence of three different mobile ions, which we attributed to $V_\text{MA}^-$, $\text{MA}_i^+$ and $\text{I}_i^-$, respectively. We introduced an improved numerical implementation of regularization algorithm RegSLapS for performing the inverse Laplace transform. Our algorithm is a powerful instrument for the evaluation of any kind of exponential decays and is especially useful to extract components of DLTS transients. By using RegSLapS we found that all measured mobile ions correspond to distribution of diffusion constants. This result could be the explanation for the different ionic defect  parameters  as reported in literature. A range of different DLTS modes as well as impedance spectroscopy verifies our findings.

Future developments in the field of perovskite solar cells depend on a detailed understanding of the ionic transport mechanisms and the degradation pathways influenced by ionic species. Our method and experimental results open routes for improvements in the fabrication process of perovskite solar cells in order to decrease the density of mobile ions within the active layer. Since the defect landscape may depend on the sample preparation conditions, a continuation of this study should include also the dependence of the defect parameters on the fabrication parameters such as the precursor solution stoichiometry.

\begin{acknowledgments}

C.D.\ and S.R.\ acknowledge financial support by the Bundesministerium für Bildung und Forschung (BMBF Hyper project, contract no.\ 03SF0514C) and thank their project partners from the University of Würzburg and ZAE Bayern for interesting discussions. Y.V.\ and C.D.\ thank the DFG for generous support within the framework of SPP 2196 project (PERFECT PVs). The authors thank Aron Walsh for his feedback concerning the interpretation of the ionic species.

\end{acknowledgments}

\end{document}

\title{Supporting Information \\
Improved evaluation of deep-level transient spectroscopy on perovskite solar cells reveals ionic defect distribution}

\author{Sebastian Reichert}
\affiliation{Institut f\"ur Physik, Technische Universit\"at Chemnitz, 09126 Chemnitz, Germany}
\author{Jens Flemming}
\affiliation{Fakult\"at f\"ur Mathematik, Technische Universit\"at Chemnitz, 09126 Chemnitz, Germany}
\author{Qingzhi An}
\author{Yana Vaynzof}
\affiliation{Kirchhoff-Institut f\"ur Physik and Centre for Advanced Materials, Ruprecht-Karls-Universit\"at Heidelberg, Im Neuenheimer Feld 227, 69120 Heidelberg, Germany}
\affiliation{Technische Universität Dresden, Institut für Angewandte Physik and Centre for Advancing Electronics Dresden (cfaed), Nöthnitzer Straße 61, 01069 Dresden}
\author{Jan-Frederik Pietschmann}
\affiliation{Fakult\"at f\"ur Mathematik, Technische Universit\"at Chemnitz, 09126 Chemnitz, Germany}
\author{Carsten Deibel}
\affiliation{Institut f\"ur Physik, Technische Universit\"at Chemnitz, 09126 Chemnitz, Germany}

\date{\today}

\maketitle

\setcounter{equation}{0}
\setcounter{figure}{0}
\setcounter{table}{0}
\setcounter{page}{1}
\setcounter{section}{0}
\makeatletter
\renewcommand{\theequation}{S\arabic{equation}}
\renewcommand{\thefigure}{S\arabic{figure}}
\renewcommand{\bibnumfmt}[1]{[S#1]}
\renewcommand{\citenumfont}[1]{S#1}
%%%%%%%%%% Prefix a "S" to all equations, figures, tables and reset the counter %%%%%%%%%%

\newpage

\section{RegSLapS: Test of validity and resolution with synthetic transients}

\begin{figure}[h]
  \includegraphics*[scale=0.7]{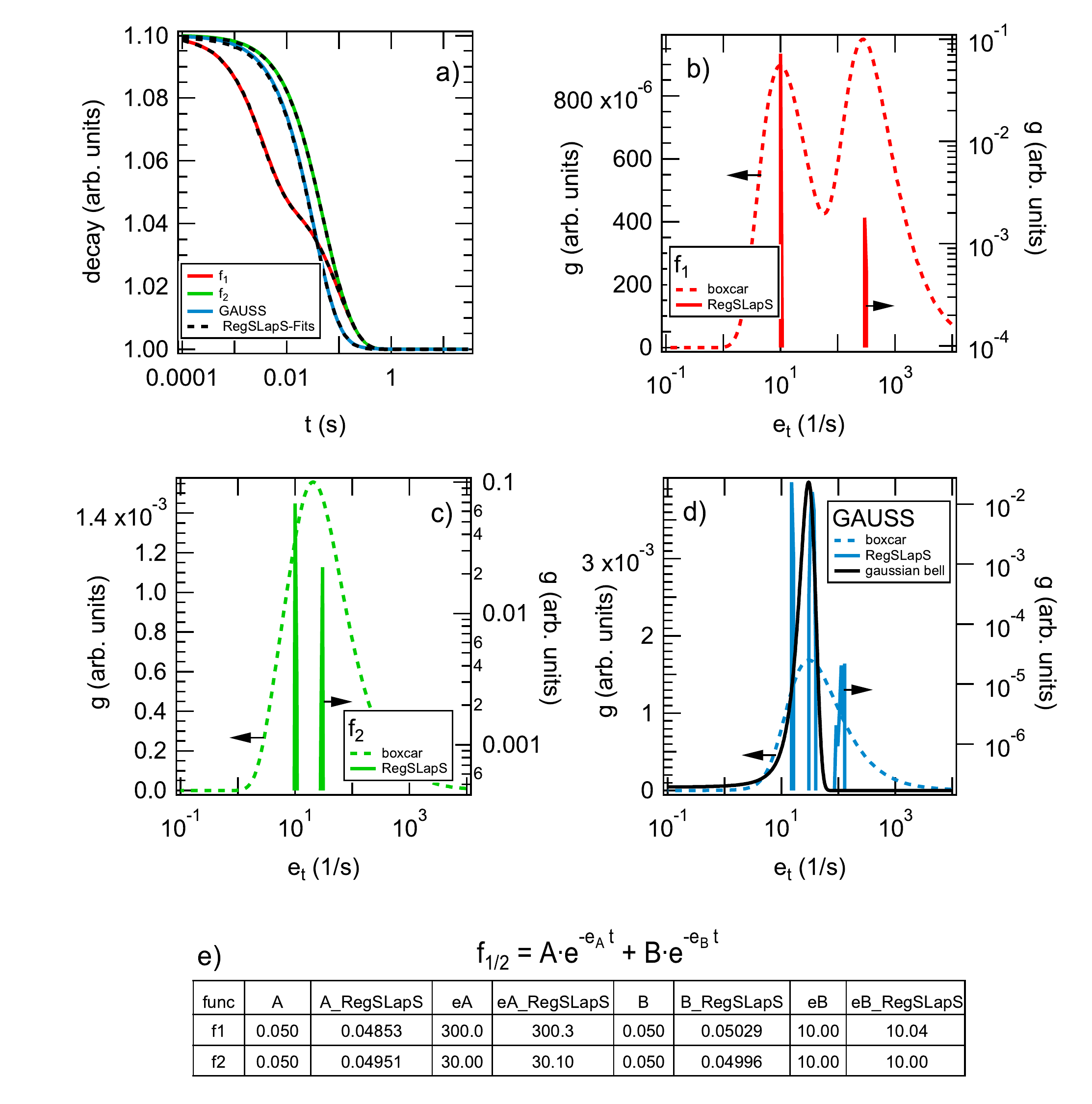}
  \caption{a) Transients of three different multi-exponential decays (see e) for parameter of f\textsubscript{1} and f\textsubscript{2}, see d) for Gaussian distribution). The ratio of the rates in the multi-exponential decays was chosen to 30 for f\textsubscript{1} and 3 for f\textsubscript{2} to highlight the gain of time resolution of RegSLapS in comparison to conventional boxcar evolution. As shown in the spectra of b) and c), boxcar evaluation is not able to resolve the two peaks in case of f\textsubscript{2}. As summarized in e) RegSLapS is able to reproduce exactly the given parameters of the multi-exponential decays. The spectra of d) elucidates that equidistant peaks are obtained by RegSLapS instead of the given Gaussian distribution.}
  \label{fig:test}
\end{figure}

\newpage

\section{Current density--voltage (jV) characteristics}

\begin{figure}[h]
  \includegraphics*[scale=0.55]{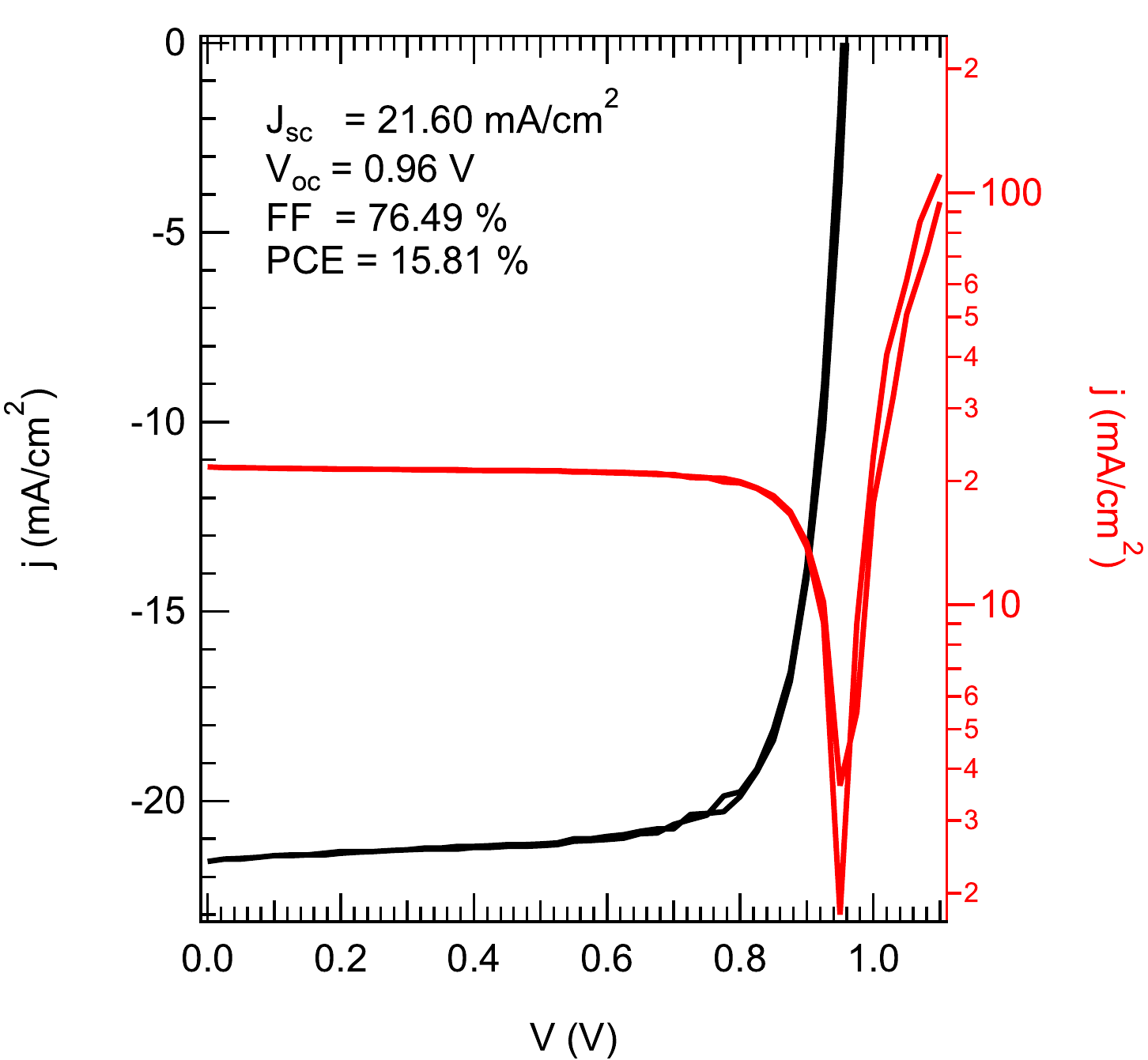}
  \caption{Current density--voltage (jV) measurement of the MAPbI\textsubscript{3} solar cell. The jV characteristics are plotted inside the graph: ${J_\mathrm{sc}=21.60~\mathrm{mA/cm^2}}$, ${V_\mathrm{oc}= 0.96~\mathrm{V}}$, ${FF=76.49~\%}$ and ${PCE=15.81~\%}$. Minimal hysteresis is visible in measurement which is caused by mobile ions within the MAPbI\textsubscript{3} layer.}
  \label{fig:IV}
\end{figure}

\section{DLTS measurements evaluation by boxcar}

\begin{figure}[h]
  \includegraphics*[scale=0.55]{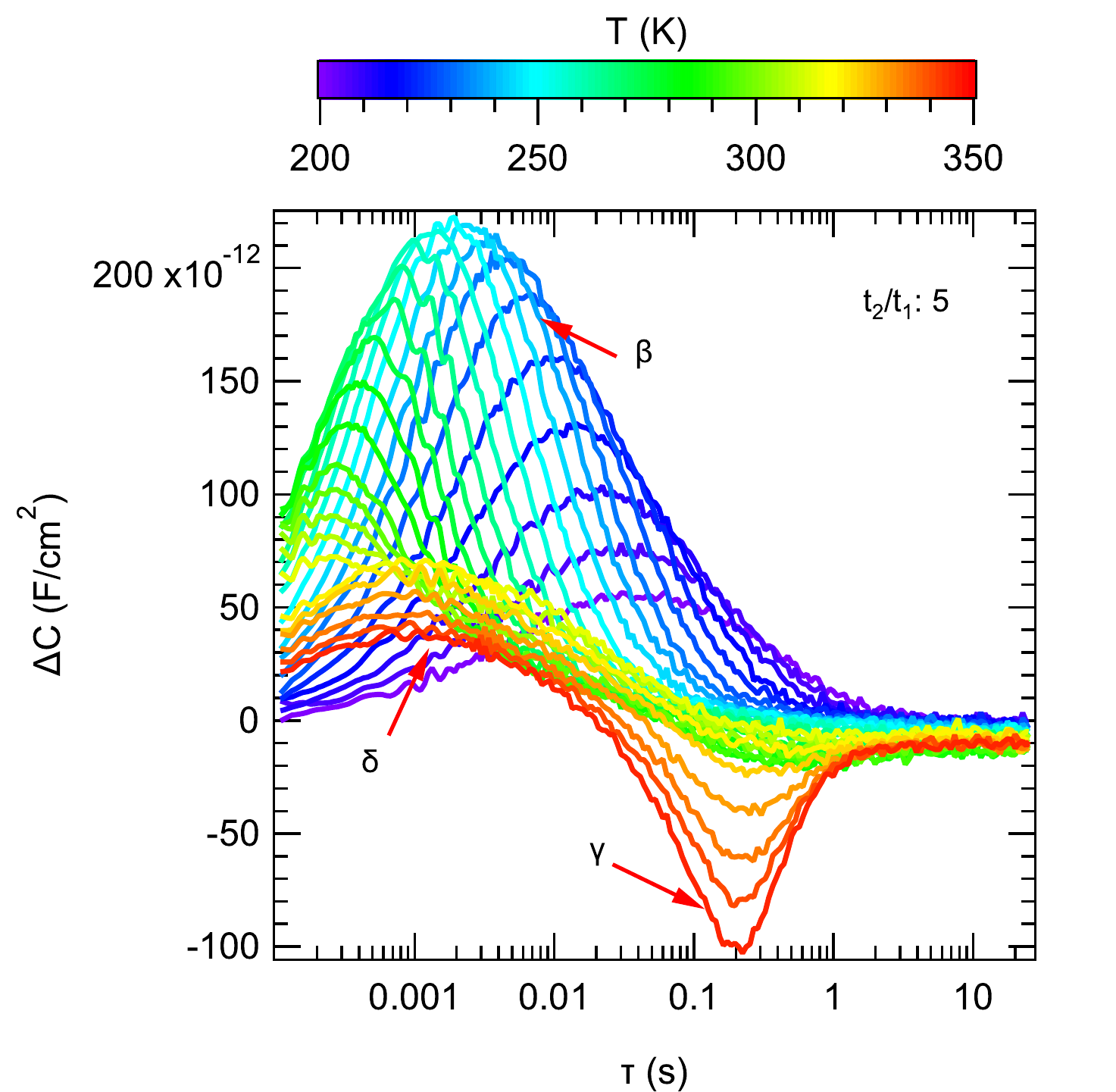}
  \caption{Rate window evaluation for the DLTS transients for a rate window of $t_2/t_1=5$. Visible are three different ionic responses, which were labeled with $\beta$, $\gamma$ and $\delta$.}
  \label{fig:boxcar}
\end{figure}

\newpage

\section{Capacitance-voltage measurements}

\begin{figure}[h]
  \includegraphics*[scale=0.65]{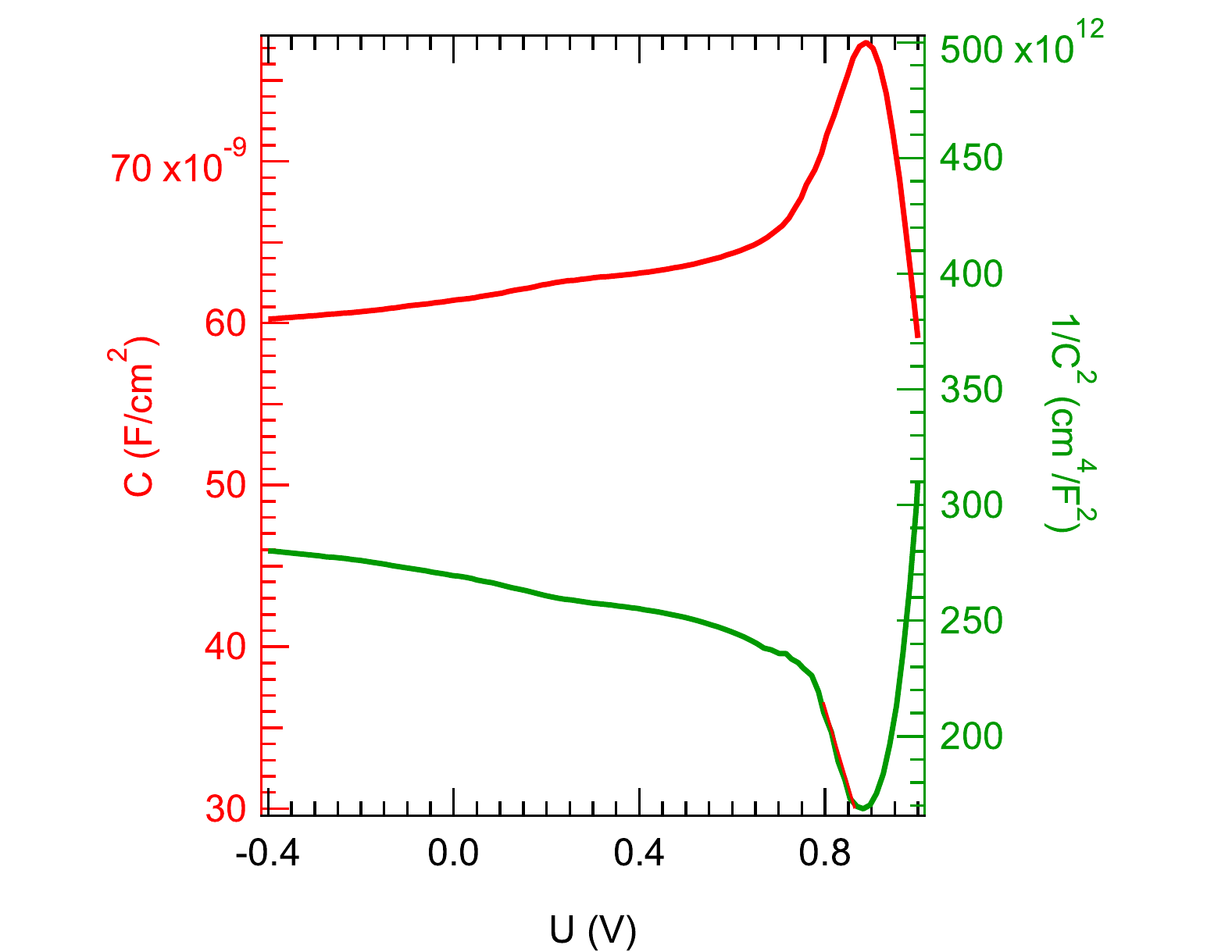}
  \caption{Capacitance-voltage (CV) measurements (left axis) and Mott-Schottky representation (right axis) at room temperature. Built-in potential $V_\mathrm{bi}$, effective doping density $N_\mathrm{eff}$ are obtained by a linear fit at around 0.8V in the Mott-Schottky plot as follows: $V_\mathrm{bi}=1.13~\mathrm{V}$ and $N_\mathrm{eff}=1.8\cdot10^{16}~\mathrm{cm^{-3}}$. The permittivity of the cell is 20.3. For accurate extraction of these parameters, the solar cells was pre-biased at 1~V for 60~s and rapidly swept with 3~V/s in reverse direction.\cite{Fischer2018}}
  \label{fig:CV}
\end{figure}

\section{Comparison of emission rates with Futscher et al.}

\begin{figure}[h]
  \includegraphics*[scale=0.65]{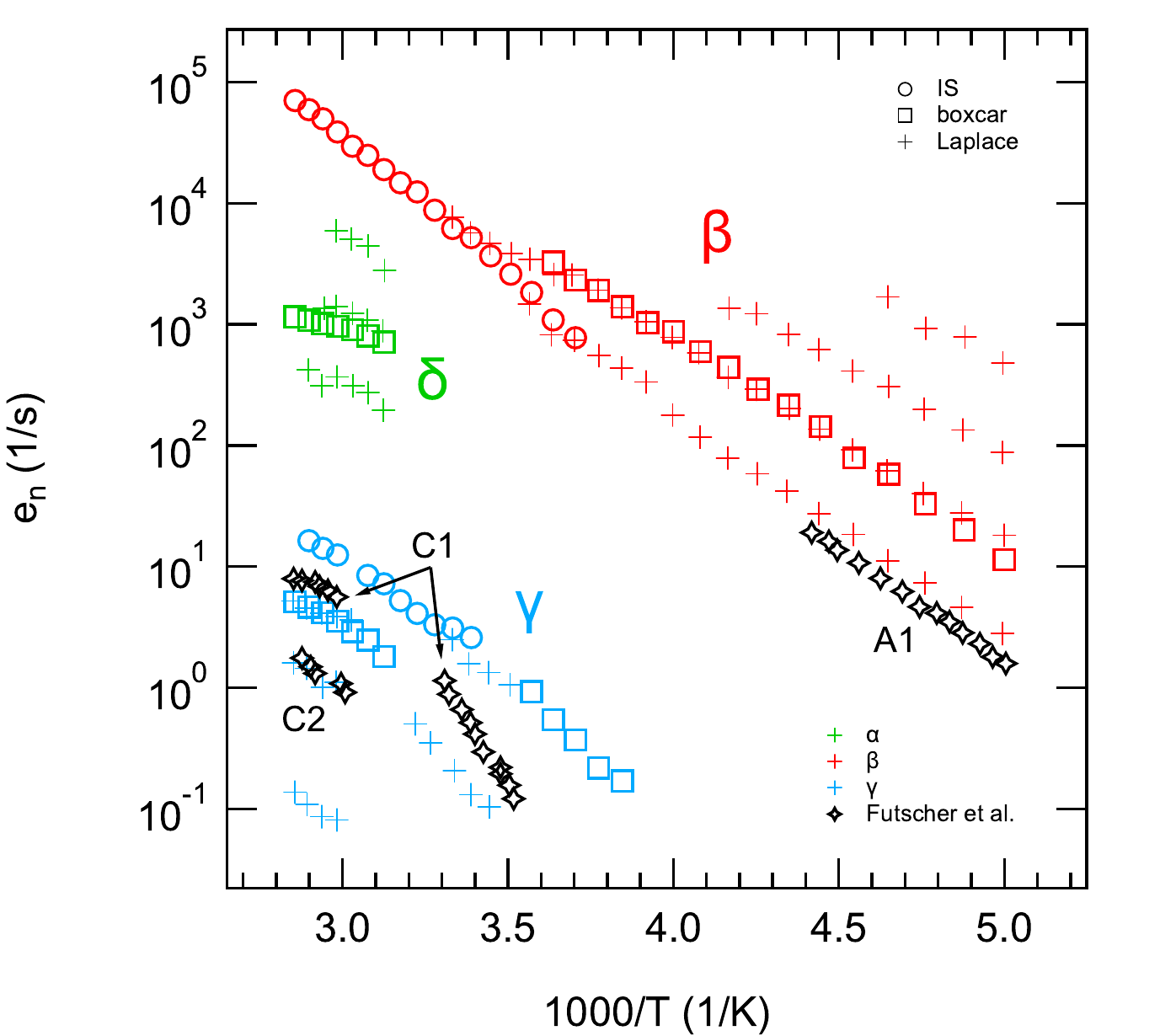}
  \caption{Emission rates of $\beta$, $\gamma$ and $\delta$ in comparison with results of \textcite{Futscher2019}. Defect $\gamma$ can be associated with defects C1 and C2, whereas all responses stem from the same defect distribution. Defect A1 is close to beta, but it can not be excluded, that A1 is a low frequency part defect $\delta$. For the identification of $\beta$ and $\gamma$, further comparisons with literature are necessary.}
  \label{fig:Futscher}
\end{figure}

\newpage

\section{Arrhenius Plot with raw data obtained by RegSLapS}

\begin{figure}[h]
  \includegraphics*[scale=0.5]{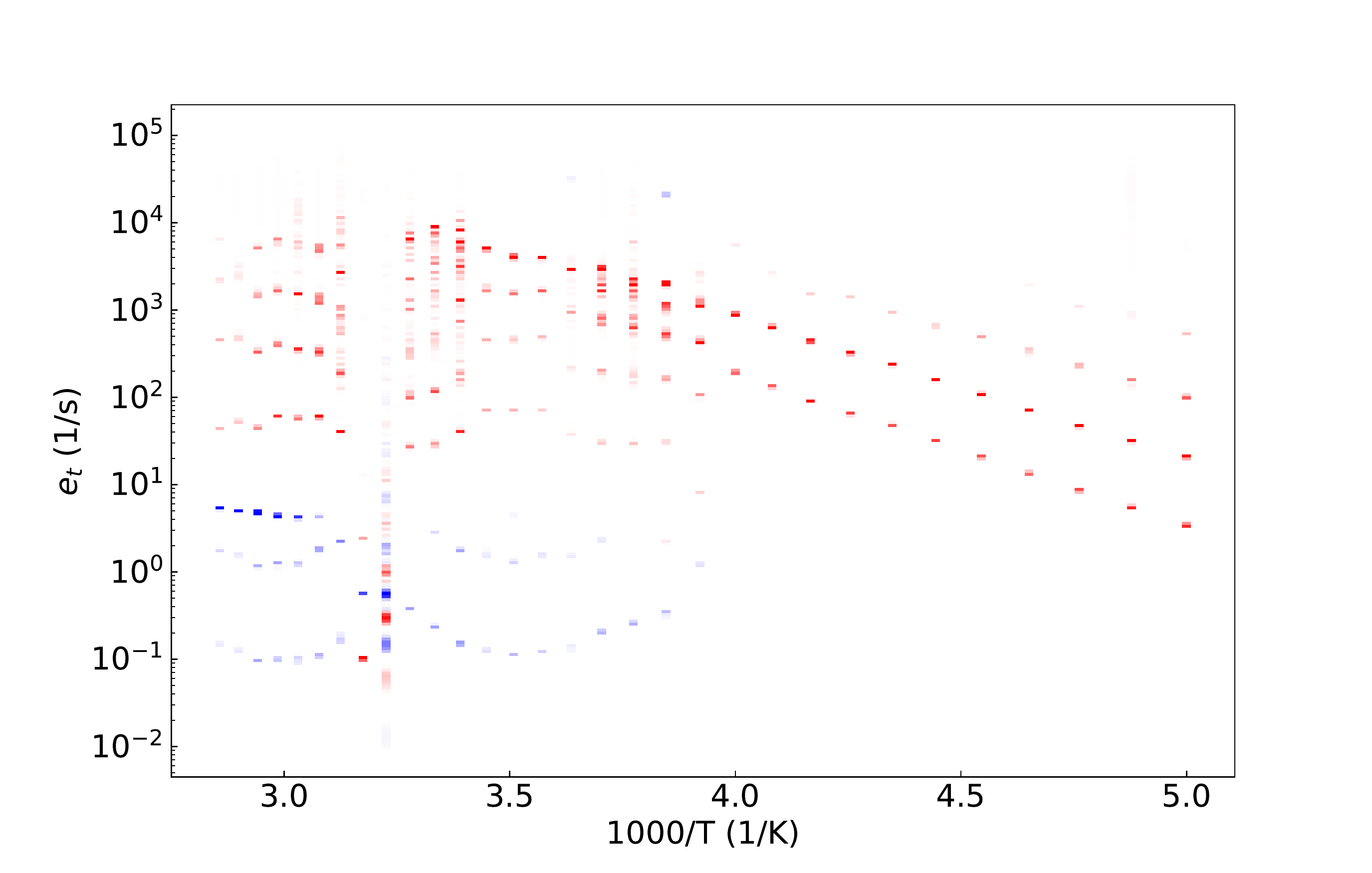}
  \caption{Emission rates computed by inverse Laplace transform using RegSLapS. The color of the points in the spectrum correspond to the sign of the defect contribution to the spectrum (blue: majority defect, red: minority defect). The transparency of the color provides information about peak height, whereas the width of the data points are connected to the width of each peak. Inverse Laplace calculation is an ill-posed problem which can cause artefacts in the spectrum. Therefore only emission rates are physically interesting, which follows the emission rate equation.}
  \label{fig:CDLTS_raw}
\end{figure}

\section{Impedance Spectroscopy (IS) measurements}

\begin{figure}[h]
  \includegraphics*[scale=0.5]{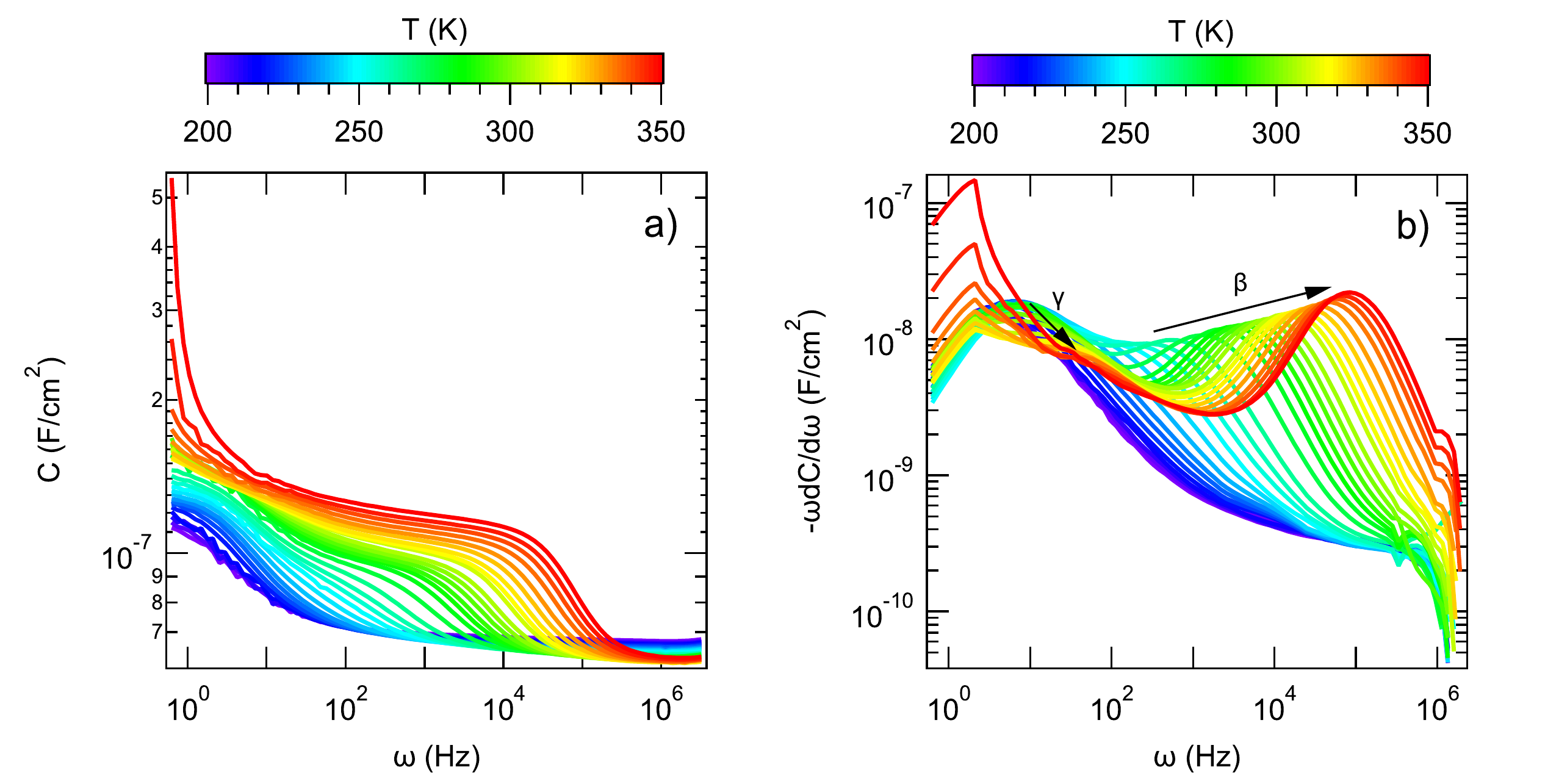}
  \caption{Impedance spectroscopy (IS) on MAPbI\textsubscript{3} solar cells for different temperatures from 200~K to 350~K in 5~K steps (a). Evaluation of the IS data by calculating the derivation (b). Visible are the defects $\beta$ and $\gamma$. Peaks of the derivation are related to the emission rates at each temperature.}
  \label{fig:IS}
\end{figure}

\newpage

\section{O-DLTS measurements}

\begin{figure}[h]
  \includegraphics*[scale=0.65]{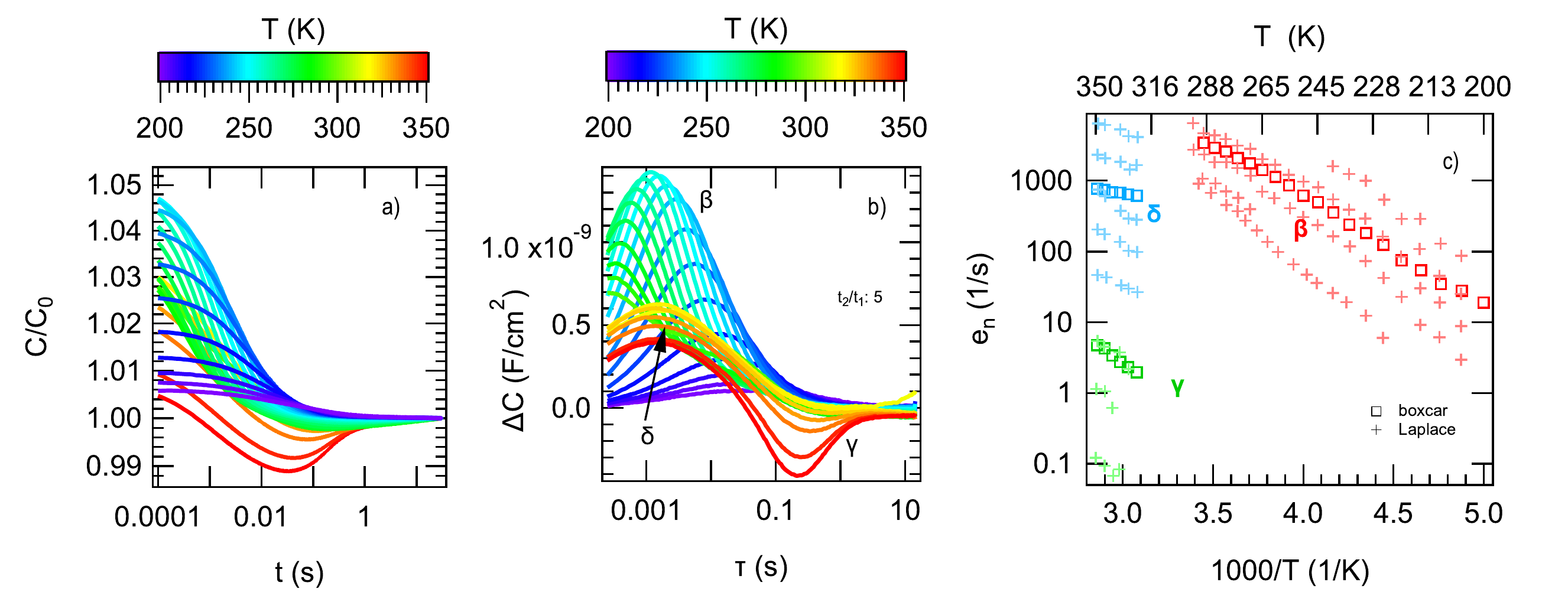}
  \caption{a) O-DLTS measurements using an light filling pulse with intensity of 1 sun for 100~ms at an AC frequency of 80~kHz with amplitude of $V_\mathrm{ac}=20~\mathrm{mV}$. b) Boxcar evaluation of the transients reveals the same ionic defect distributions measured with C- and I-DLTS. c) Emission rates obtained by boxcar evaluation (b)) and by performing inverse Laplace transformation using RegSLapS.}
  \label{fig:ODLTS}
\end{figure}

\section{I-DLTS measurements}

\begin{figure}[h]
  \includegraphics*[scale=0.65]{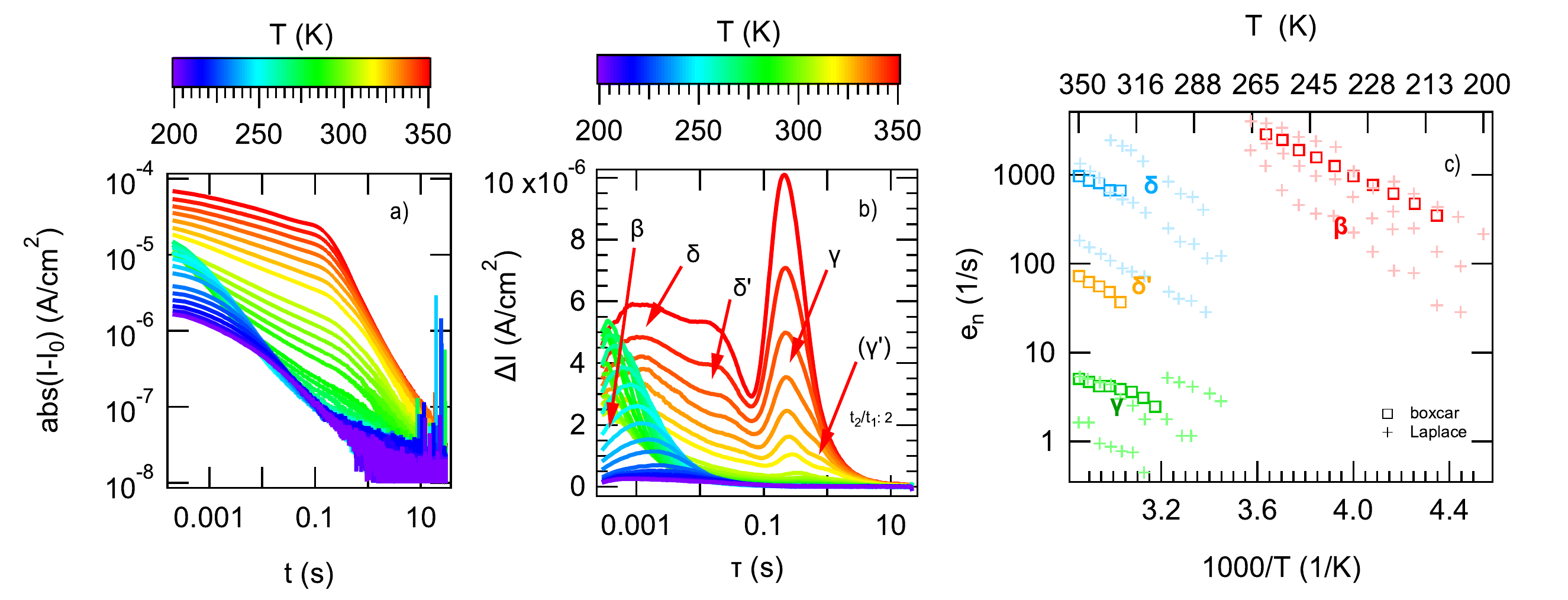}
  \caption{a) I-DLTS transients were measured choosing an AC frequency of 80~kHz with amplitude of $V_\mathrm{ac}=20~\mathrm{mV}$ and using a filling pulse from 0~V to 1~V for 100~ms. Measurements were averaged over 35 single measurements and were measured simultaneously with C-DLTS transients. Boxcar evaluation of the transients reveals the same ionic defect distributions measured with C- and O-DLTS. The higher resolution of I-DLTS especially at high emission rates gives insight into different parts of the distributions. c) Emission rates obtained by boxcar evaluation (b)) and by performing inverse Laplace transformation using RegSLapS.}
  \label{fig:IDLTS}
\end{figure}

\newpage

\section{Comparison of C-, I- and O-DLTS measurements}

\begin{figure}[h]
    \includegraphics*[scale=0.60]{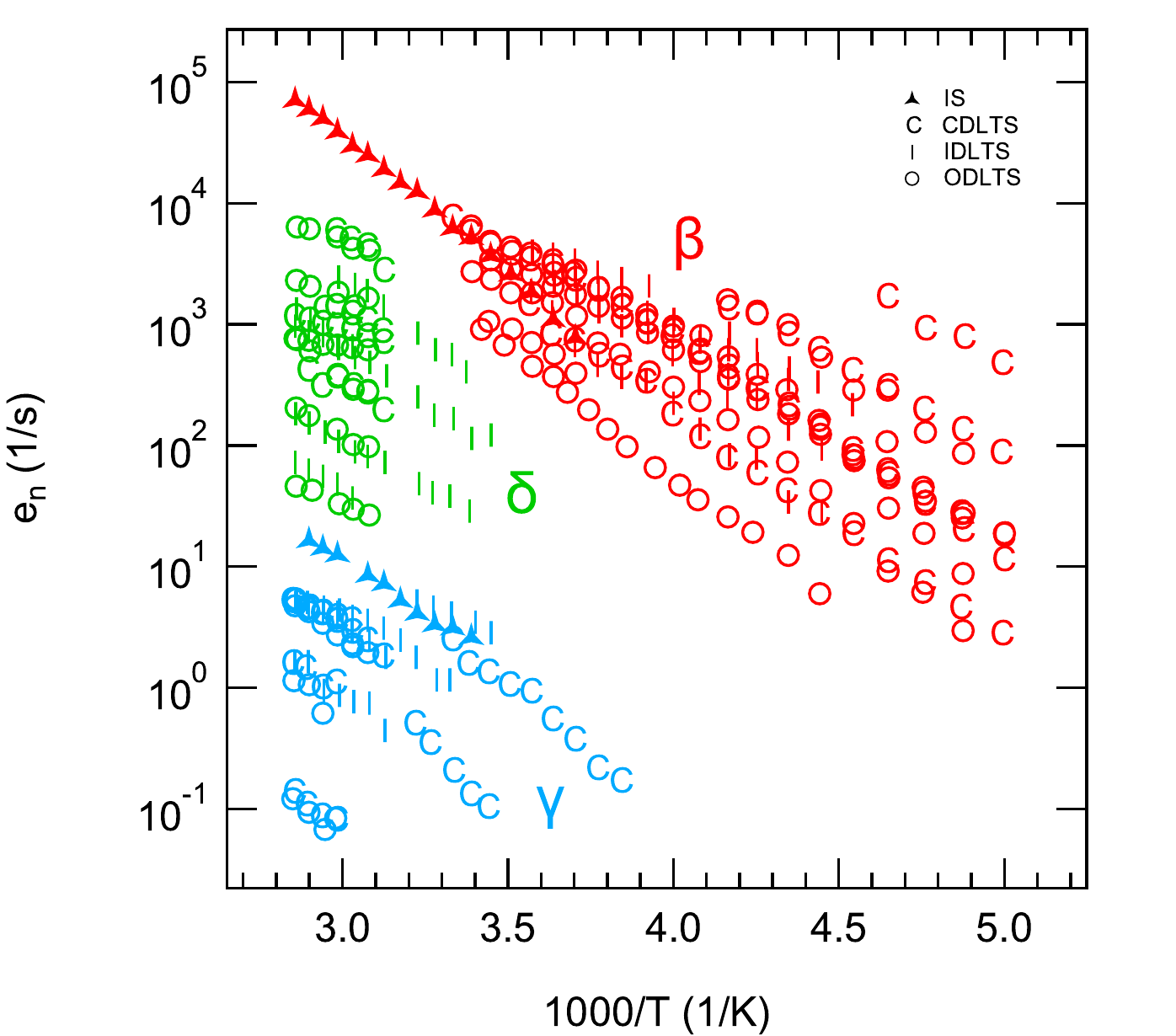}
    \caption{Comparison of emission rates of the ionic defects $\beta$, $\gamma$ and $\delta$ obtained by C-, I-, O-DLTS (boxcar and Laplace evaluation) and IS (Fig.~\ref{fig:IS}). With all DLTS modes the same defect distributions are measurable. The plot illustrate how close the distributions are spaced.}
    \label{fig:CIO}
\end{figure}